\documentclass[12pt]{article}

\usepackage{fullpage,color,ulem}
\usepackage{natbib}
\usepackage{url}
\usepackage{placeins}
\usepackage{hyperref}
\usepackage{amsmath,amssymb}
\usepackage{graphicx}
\usepackage[symbol]{footmisc}
\begin{document}

\begin{center}

{\LARGE \texttt{phylodyn}: an R package for phylodynamic simulation and inference}\\\ \\
{ Michael D. Karcher\textsuperscript{1*}, Julia A. Palacios\textsuperscript{2,3}\footnote{The first two authors contributed equally to this paper.},
  Shiwei Lan\textsuperscript{4}, Vladimir N. Minin\textsuperscript{1,5}}\\
  {\footnotesize \textsuperscript{1}Department of Statistics, University of Washington, Seattle, WA, USA\\
  \textsuperscript{2}Department of Statistics, Stanford University, Stanford, CA, USA \\
  \textsuperscript{3}Department of Biomedical Data Science, Stanford University, Stanford, CA, USA\\
  \textsuperscript{4}Department of Statistics, University of Warwick, Coventry, UK\\
  \textsuperscript{5}Department of Biology, University of Washington, Seattle, WA, USA\\
  }
\end{center}

\begin{abstract}
We introduce \texttt{phylodyn}, an R package for phylodynamic analysis based on gene genealogies. 
The package main functionality is Bayesian nonparametric estimation of effective population size fluctuations over time. 
Our implementation includes several Markov chain Monte Carlo-based methods and an integrated nested Laplace approximation-based approach for phylodynamic inference that have been developed in recent years.
Genealogical data  describe the timed ancestral relationships of individuals sampled from a population of interest. Here, individuals are assumed to be sampled at the same point in time (isochronous sampling) or at different points in time (heterochronous sampling); in addition, sampling events can be modeled with preferential sampling, which means that the intensity of sampling events is allowed to depend on the effective population size trajectory. 
We assume the coalescent and the sequentially Markov coalescent  processes as generative models of genealogies. 
We include several coalescent simulation functions that are useful for testing our phylodynamics methods via simulation studies. 
We compare the performance and outputs of various methods implemented in \texttt{phylodyn} and outline their strengths and weaknesses.
R package \texttt{phylodyn} is available at \url{https://github.com/mdkarcher/phylodyn}.
\end{abstract}

\section*{Introduction}

In the last several decades, phylodynamic inference has demonstrated its usefulness in ecology and epidemiology \citep{grenfell2004unifying, holmes2009discovering}.
The key inferential insight of phylodynamics is that population dynamics leave their mark in the shape of gene 
genealogies and thereby the sequence data sampled.
Kingman's coalescent models the relationship between effective population size $N_{e}(t)$
and the likelihood of observing a particular genealogy \citep{coalescent}.
In order to be computationally feasible, early coalescent-based models required
strong parametric assumptions on the effective population size trajectory
\citep{griffiths1994sampling, drummond2002estimating, kuhner1998maximum}.
More recently, nonparametric models have allowed a much more diverse class of effective
population size trajectories to be inferred, at the cost of estimating many more parameters.
Methods have emerged that compromise between the two extremes, maintaining a tractable
number of parameters while allowing for a diverse class of estimable trajectories
\citep{drummond2005bayesian, skyride,palacios2013gaussian, skygrid}.
See the review by \citet{ho2011skyline} for a detailed comparison.

Here we unify user interfaces for three different but related Bayesian nonparametric methods. These methods assume a log Gaussian process prior on $N_{e}(t)$.
The first comes from the work by \citet{lan2015efficient}.
They implement a number of Markov chain Monte Carlo (MCMC) algorithms for inferring
effective population size trajectories from a fixed genealogy.
They compare different algorithms' computational efficiency and MCMC diagnostics.

The second methodology comes from the work by \citet{palacios2012INLA} and \citet{karcher2015quantifying}.
They target the same posterior as in \citep{lan2015efficient}, but implement
an integrated nested Laplace approximation (INLA) based approach.
Utilizing INLA allows for a significant computational speedup at the cost of
only having access to the latent parameters' approximate marginal distributions
(as opposed to MCMC algorithms which approximate the full joint distribution).
\citet{karcher2015quantifying} have an additional focus of accounting for potential preferential sampling, which incorporates a likelihood relating
the sampling times of the genealogy to the effective population size trajectory.
\par
The last methodology comes from the work by \citet{Palaciosgenetics}. They implement an MCMC algorithm for inferring effective population size trajectories from a sequence of local genealogies. Here, genealogies are correlated and are assumed to be a realization of the sequentially Markov coalescent (SMC$'$) \citep{MarjoramSMC}.
\par
An R package \texttt{phylodyn} encapsulates all the above work. 
We integrated all of the above methods in a unified user-friendly format, added detailed tutorials, included more features such as simulation of genealogies from the coalescent model that accepts arbitrary but positive effective population size function \citep{palacios2013gaussian}, and added features for data manipulation and interaction with other data formats such as BEAST-XML \citep{BEAST}. 
These features greatly expand available phylodynamics methods in \texttt{R}.
For example, the \texttt{R} package \texttt{ape} \citep{ape} has a function \texttt{skyline} that implements the generalized skyline method for isochronous genealogies. To the best of our knowledge, no other \texttt{R} package infers effective population size trajectories from heterochronous genealogies. Other \texttt{R} packages for simulation of genealogical data exist (e.g.\ \texttt{phyclust} \citep{phyclust} and \texttt{ape}) but they are limited to very specific demographic scenarios such as piece-wise constant and exponential growth functions.
Our addition of inference from a sequence of local genealogies expands \texttt{phylodyn} to a broader class of models that have not been implemented in the previous versions of the package.


\section*{Functionality}

\subsection*{Genealogical simulation}

A genealogy is a rooted bifurcating tree with labeled tips. Branching events are called coalescent events which occur at coalescent times, and tips are located at sampling times. Given a vector of sampling times $\mathbf{s}$ and an effective population size function
$N_e(t)$, Kingman's coalescent provides the following likelihood of observing a particular genealogy $\mathbf{g}$ with coalescent times $\mathbf{t} = \left\{ t_i \right\}_{i=2}^n$:
\[
\Pr[\mathbf{g} | N_e(t), \mathbf{s}] \propto \prod_{k=2}^{n} {\frac{C_{0,k}}{N_e(t_{k-1})}} \exp\left[ -\sum_{i=0}^{m_k} {\int_{I_{i,k}} {\frac{C_{i,k}}{N_e(t)} dt}} \right],
\]
where $C_{i,k}=\binom{n_{i,k}}{2}$, $n_{i,k}$ is the number of lineages present during time interval $I_{i,k}$, and $I_{i,k}$ is a time interval defined by coalescent times and sampling times and $I_{0,k}$ is a time interval that ends at coalescent time $t_{k-1}$. See \citep{lan2015efficient} for notational details. The \texttt{coalsim} function simulates coalescent times according to this
distribution, given a vector of sampling times and an arbitrary effective population size function \texttt{traj(t)}.
The function gives the option of using a time-transformation method or
a thinning method for simulating the coalescent times.
The time-transformation method scales better but involves numerical integration,
while the thinning method is faster with few samples and is an exact method.  The \texttt{generate\_newick} function takes the output  generated with \texttt{coalsim} and returns the corresponding genealogy in \texttt{ape}'s phylo format \citep{ape}.
We are not aware of another R package that allows for simulating the coalescent process while allowing for arbitrary sampling times as well as arbitrary positive effective population size trajectories.
\texttt{phylodyn} also provides functionality for easily simulating sampling times
under preferential sampling according to an arbitrary positive function $f$.
The \texttt{pref\_sample} function simulates sampling times according to an
inhomogeneous Poisson process with intensity
$\lambda(t) = c f(t)^{\beta}$, where parameters $c$ and $\beta$
control the  expected number of 
sampled sequences and the strength of preferential sampling, respectively.
Currently the function only allows a thinning method, but a time-transformation
method is forthcoming.

\subsection*{Markov chain Monte Carlo methods}

Following the approach of \citet{skygrid} and \citet{palacios2012INLA}, \citet{lan2015efficient} approximate $N_{e}(t)$ by a piece-wise linear function 
$N_f(t) = \sum_{d=1}^{D-1} {\exp(f_d) 1_{\left(x_d, x_{d+1}\right]}}$, defined over a regular grid with end points $\mathbf{x} = \left\lbrace x_d \right\rbrace_{d=1}^D$, 
where $x_1$ equals the most recent sampling time,
and $x_D=t_{2}$, the time when the last two lineages coalesce. Hence, we seek to estimate the posterior
\begin{equation}
\Pr[\mathbf{f}, \tau \mid \mathbf{g}] \propto \Pr[\mathbf{g} \mid \mathbf{f}] \Pr[\mathbf{f} \mid \tau] \Pr(\tau), \label{posterior}
\end{equation}
where $\Pr[\mathbf{g} \mid \mathbf{f}]$ is the coalescent likelihood, $\Pr[\mathbf{f} \mid \tau]$ is a Gaussian process prior on $\mathbf{f} = \left\lbrace f_d \right\rbrace_{d=1}^{D-1}$ with precision $\tau$, and $\Pr(\tau)$ is a Gamma hyperprior on $\tau$. Our implementation assumes a Gaussian process prior on $\mathbf{f}$ with inverse covariance function $\mathbf{C}^{-1}(\tau)=\frac{1}{\tau}\mathbf{C}^{-1}$, where $\mathbf{C}^{-1}$ corresponds to a modified inverse covariance matrix of Brownian motion (see \citep{lan2015efficient} for details).

The \texttt{mcmc\_sampling} function implements a variety of MCMC algorithms for
estimating the posterior \eqref{posterior},
given the sufficient statistics for a genealogy (sampling times and coalescent times).
Available methods are Hamiltonian Monte Carlo (HMC) \citep{duane1987hybrid, neal2011mcmc},
split HMC \citep{leimkuhler2004simulating, neal2011mcmc, shahbaba2014split},
Metropolis-adjusted Langevin algorithm (MALA) \citep{roberts1996exponential},
adaptive MALA \citep{knorr2002block},
and Elliptical Slice Sampler (ESS) \citep{murray2010elliptical}. For a comparison of the computational efficiency of the different methods
see \citep{lan2015efficient}.

\begin{figure}[htbp]
	\centering
	\includegraphics[width=1.0\textwidth]{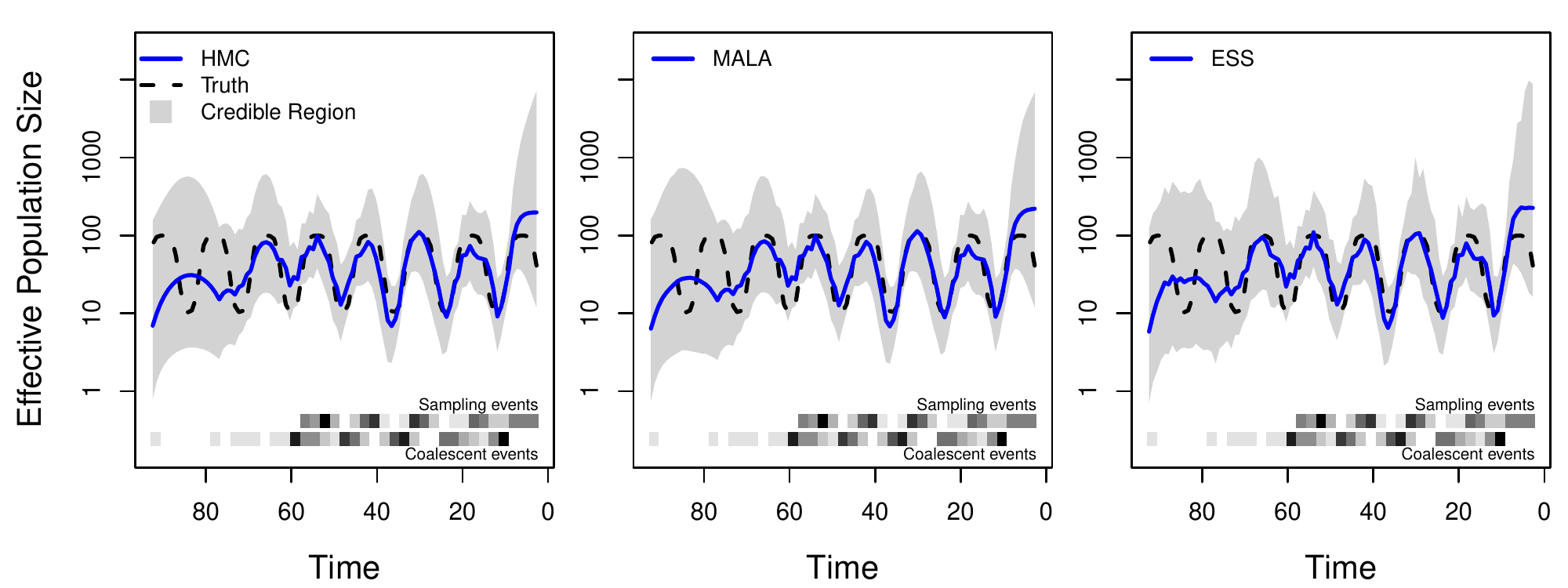}
	\caption{ Seasonal boom-and-bust population size trajectory recovered with three different MCMC estimation methods: HMC, MALA and ESS.  
		The dashed black lines represent the true population size trajectory.
		The solid blue lines represent the posterior median estimates, and the shaded regions represent the 95\% credible regions.
		At bottom, the upper and lower heatmaps represent freGenequencies of sampling events and coalescent events, respectively.
		Time in simulated units of weeks.
		}
	\label{fig:MCMC}
\end{figure}

We illustrate \texttt{phylodyn}'s capabilities with a simulation example.
We let $N_e(t)$ have a seasonal boom-and-bust trajectory
(provided by the \texttt{logistic\_traj} function),
and we simulate a sequence of sampling times according to an
inhomogeneous Poisson process with intensity proportional to $N_e(t)$
using the \texttt{pref\_sample} function.
We simulate a genealogy from the coalescent using the \texttt{coalsim} function,
and supply it to the different sampling algorithms of the \texttt{mcmc\_sampling} function.
We summarize the results in Figure \ref{fig:MCMC}.

\begin{figure}[htbp]
	\centering
	\vspace{-0.6cm}
	\includegraphics[width=0.9\textwidth]{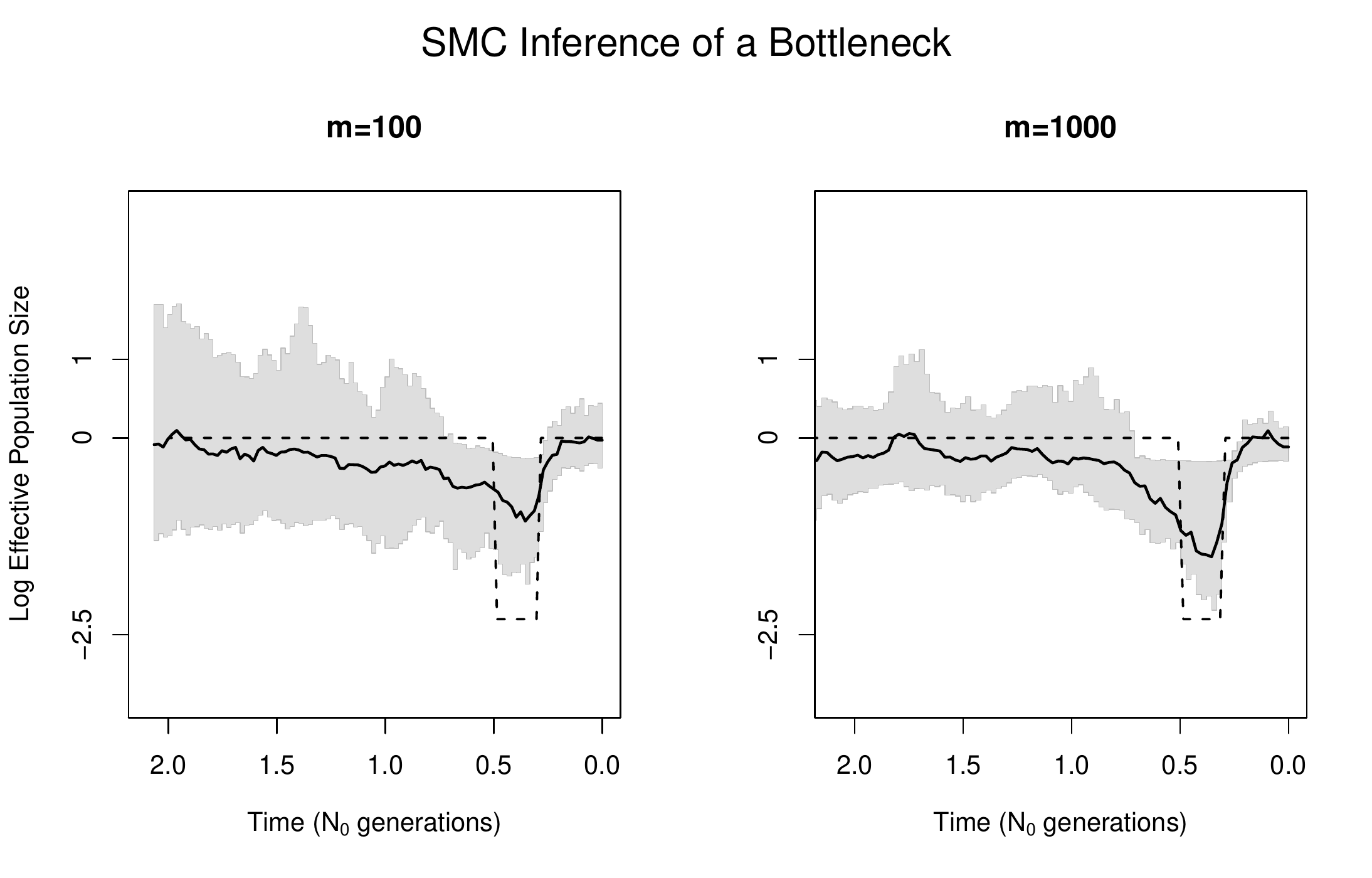}
	\vspace{-0.8cm}
	\caption{ SMC' inference of $N_{e}(t)$ from $m=100$ and $m=1000$ simulated local genealogies of $n=20$ individuals.
		The dashed black line represents the true population size trajectory, the solid black line represents the posterior median estimates, and the shaded regions represent the 95\% credible regions. Estimation improves with larger number of genealogies.}
	\label{fig:smc}
\end{figure}

\citet{Palaciosgenetics} infer $N_{e}(t)$ from a sequence of $m$ local genealogies under the SMC' model \citep{MarjoramSMC}. 
The SMC' process is an approximation to the ancestral recombination graph (ARG) which models the set of ancestral relationships and recombination events of multilocus sequences \citep{griffiths1997ancestral}. In our implementation, we assume that our data consist of a sequence of genealogies that represent the ancestral relationships at consecutive loci separated by recombination events. These consecutive genealogies are modeled as a continuous-time Markov chain along a chromosomal segment. Here, we also approximate $N_{e}(t)$ by the piece-wise linear function $N_{f}(t)$ and rely on split HMC  \citep{shahbaba2014split} to sample from the posterior:
\begin{equation}\label{eq:poster}
\text{Pr}[\mathbf{f},\tau\mid \mathbf{g}_{0},\ldots,\mathbf{g}_{m-1}] \propto \text{Pr}[\mathbf{g}_{0}\mid \mathbf{f}]
\times \left\lbrace \prod^{m-2}_{i=0}\text{Pr}[\mathbf{g}_{i+1} \mid \mathbf{g}_{i}, \mathbf{f}] \right\rbrace \text{Pr}[\mathbf{f} \mid \tau] \text{Pr}(\tau),
\end{equation}
where $\text{Pr}[\mathbf{g}_{0},\ldots,\mathbf{g}_{m-1}\mid \mathbf{f}]$ is the sequentially Markov coalescent likelihood \citep{Palaciosgenetics}. Our \texttt{mcmc\_smc} function samples from the posterior distribution (\ref{eq:poster}). Figure \ref{fig:smc} shows our estimate of $N_{e}(t)$ from 100 and 1000 local genealogies  of $n=20$ individuals simulated under a bottleneck demographic scenario. \citet{Palaciosgenetics} show that our method recovers the bottleneck best when increasing the number of local genealogies. 
 
\subsection*{INLA-based methods}

We implement the INLA-based methods of \citet{palacios2012INLA} and \citet{karcher2015quantifying}, using the same log-Gaussian prior on $N_{e}(t)$ as in the MCMC methods. 
The \texttt{BNPR} function implements the INLA approximation to obtain posterior medians and 95\% Bayesian credible intervals (BCIs) of $N_{f}(t)$.
Being a numerical approximation, 
this method runs extremely quickly. However, the method only estimates the marginals of the posterior of the effective population size and hyperparameters, rather than the full joint posterior distribution of MCMC-based methods. This is frequently sufficient for most purposes involving phylodynamic inference, but offers significant improvement in computational efficiency.

We also implement the BNPR-PS method of \citet{karcher2015quantifying}.
In cases where the frequency of sampling times is related to effective population size,
including a sampling time model provides additional accuracy and precision.
We model the sampling times as an inhomogeneous Poisson process
with intensity proportional to a power of the effective population size,
with the following log-likelihood:
\[
\log[\Pr(\mathbf{s}\mid \mathbf{f}, \beta_0, \beta_1)] = C + n \beta_0 + \sum_{i=1}^{n}{\beta_1 \log[N_{f}(s_i)] - \int_{s_m}^{s_0}{\exp(\beta_0) [N_f(r)]^{\beta_1} dr}}.
\]
This leads to the posterior that conditions on both coalescent and sampling times:
\begin{equation}
\Pr[\mathbf{f}, \tau, \beta_0, \beta_1 \mid \mathbf{g}, \mathbf{s}] \propto \Pr[\mathbf{g} \mid \mathbf{s}, \mathbf{f}] \Pr[\mathbf{s} \mid \mathbf{f}, \boldsymbol{\beta}] \Pr[\mathbf{f} \mid \tau] \Pr(\tau) \Pr(\beta_0, \beta_1). \label{posterior2}
\end{equation}

\begin{figure}[htbp]
	\centering
	\includegraphics[width=0.8\textwidth]{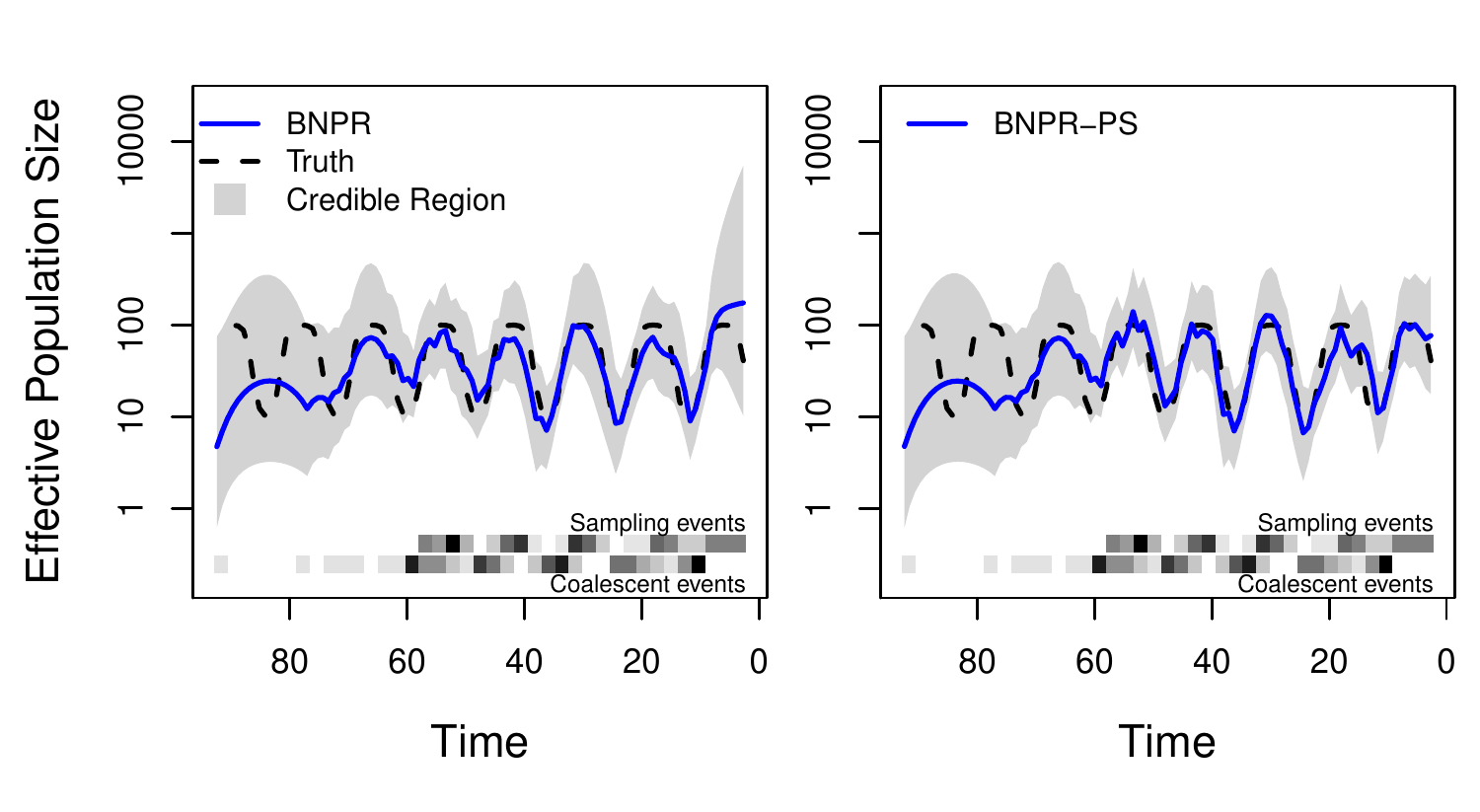}
	\caption{ Graphical representation of the output of a single genealogy simulation and different BNPR estimation methods.
		The dashed black lines represent the true population size trajectory.
		The solid blue lines represent the posterior median estimates, and the shaded regions represent the 95\% credible regions.
		The bottom upper and lower heatmaps represent frequencies of sampling events and coalescent events, respectively.
		For this figure, we sampled individuals according to an inhomogeneous Poisson process with intensity proportional to effective population size $N_e(t)$ ($\beta_1 = 1$).
		The plot on the left is generated by Bayesian nonparametric phylodynamic reconstruction (BNPR) that does not account for preferential sampling, while the plot on the right is generated by Bayesian nonparametric phylodynamic reconstruction with preferential sampling (BNPR-PS) and incorporates our sampling time model.
		Time is in months.}
	\label{fig:BNPR}
\end{figure}

To illustrate, we use the same genealogy under seasonal boom-and-bust population size trajectory as in Figure 1.
We apply BNPR and BNPR-PS to this genealogy,
and summarize the results in Figure \ref{fig:BNPR}.
Since our sampling times and genealogy were simulated with preferential sampling,
we notice improved performance from BNPR-PS,
which correctly models the sampling times.

\section*{Discussion}
Phylodynamic inference aims to enhance our understanding of infectious disease dynamics that involves a combination of evolutionary, epidemiological, and immunological processes \citep{grenfell2004unifying}.  
Although phylodynamic methods have been developed and successfully employed over the last 15 years, there are still many challenges in extending these methods to incorporate different types of information  and evolutionary complexities of certain pathogens \citep{Frost:2015gu}.   
The tools developed in \texttt{phylodyn} currently concentrate on estimation of population dynamics from genealogical and sampling information --- a subset of phylodynamics problems.
Phylodynamic inference from sequence data alone is challenging because the state spaces of genealogies $\mathbf{g}$ and effective population size trajectories $N_{e}(t)$ are large. The MCMC tools implemented in \texttt{phylodyn} allow for an efficient exploration of the state space of effective population size trajectories $N_{e}(t)$ when either a single genealogy is available or multiple local sequential genealogies are available. Future implementation in \texttt{phylodyn} will involve the exploration of the joint space of genealogies, population size trajectories and other epidemiological processes. We envision that the increasing popularity of \texttt{R} will allow researchers to integrate different packages with \texttt{phylodyn}. For instance, \texttt{phylodyn} can be used in combination with the \texttt{R} package \texttt{coalescentMCMC} to account for genealogical uncertainty.  In addition, our coalescent simulation functions should be of interest to a wide range of users of the coalescent.

\FloatBarrier

\section*{Data Accessibility}

\texttt{phylodyn} is available at \url{https://github.com/mdkarcher/phylodyn}.
Installation instructions are provided in the README file.
Several vignettes have been included to walk users through the standard workflow,
as well as a number of example datasets from the papers that introduced the methods included in the R package.

\end{document}